\documentclass[letterpaper,english,aps,prl,floatfix,reprint,superscriptaddress,showpacs,twocolumn]{revtex4-1}
\usepackage[T1]{fontenc}
\usepackage[utf8]{inputenc}
\usepackage{xcolor}
\usepackage{pdfcolmk}
\usepackage{float}
\usepackage{textcomp}
\usepackage{amsmath}
\usepackage{amssymb}
\usepackage{graphicx}
\PassOptionsToPackage{normalem}{ulem}
\usepackage{ulem}

\makeatletter

\pdfpageheight\paperheight
\pdfpagewidth\paperwidth

\providecommand{\tabularnewline}{\\}
\providecolor{lyxadded}{rgb}{0,0,1}
\providecolor{lyxdeleted}{rgb}{1,0,0}

\DeclareRobustCommand{\lyxsout}[1]{\ifx\\#1\else\sout{#1}\fi}

\pdfoutput=1
\usepackage{mathrsfs}
\usepackage{bm}
\usepackage{xspace}
\usepackage{epstopdf}
\usepackage{longtable}
\usepackage{amsmath}    
\allowdisplaybreaks[4]  

\usepackage[colorlinks,linkcolor=blue,anchorcolor=blue,citecolor=blue]{hyperref}

\makeatother

\usepackage{babel}
\begin{document}
\title{Switchable in-plane anomalous Hall effect by magnetization orientation
in monolayer $\mathrm{Mn}_{3}\mathrm{Si}_{2}\mathrm{Te}_{6}$}
\author{Ding Li}
\affiliation{Anhui Key Laboratory of Low-Energy Quantum Materials and Devices,
High Magnetic Field Laboratory, HFIPS, Chinese Academy of Sciences,
Hefei, Anhui 230031, China }
\affiliation{Department of Physics, University of Science and Technology of China,
Hefei 230026, P.R. China}
\author{Maoyuan Wang}
\email{mywang@xmu.edu.cn}

\affiliation{Department of Physics, Xiamen University, Xiamen 361005, China}
\author{Dengfeng Li}
\affiliation{School of Science, Chongqing University of Posts and Telecommunications,
Chongqing, 400065, China.}
\affiliation{Institute for Advanced Sciences, Chongqing University of Posts and
Telecommunications, Chongqing, 400065, China.}
\author{Jianhui Zhou}
\email{jhzhou@hmfl.ac.cn}

\affiliation{Anhui Key Laboratory of Low-Energy Quantum Materials and Devices,
High Magnetic Field Laboratory, HFIPS, Chinese Academy of Sciences,
Hefei, Anhui 230031, China }
\date{\today}
\begin{abstract}
In-plane anomalous Hall effect (IPAHE) is an unconventional anomalous
Hall effect (AHE) with the Hall current flows in the plane spanned
by the magnetization or magnetic field and the electric field. Here,
we predict a stable two-dimensional ferromagnetic monolayer $\mathrm{Mn}_{3}\mathrm{Si}_{2}\mathrm{Te}_{6}$
with collinear ordering of Mn moments in the basal plane. Moreover,
we reveal that the monolayer $\mathrm{Mn}_{3}\mathrm{Si}_{2}\mathrm{Te}_{6}$
possesses a substantial periodic IPAHE due to the threefold rotational
symmetry, which can be switched by changing the magnetization orientation
by external magnetic fields. In addition, we briefly discuss the impacts
of moderate strains on the electronic states and AHE, which lead to
a near quantized Hall conductivity. Our work provides a potential
platform for realizing a sizable and controllable IPAHE that greatly
facilatates the application of energy-efficient spintronic devices.
\end{abstract}
\maketitle

\section{Introduction}

The conventional AHE with the orthogonal magnetization, electric field,
and Hall current in FM materials could be crucial in understanding
dissipationless quantum phenomena \citep{Nagaosa2010RMP,Xiao2010RMP}
and realizing low-power spintronic devices \citep{linXY2019NE,polshyn2020nature,wangZR2020NRM,CaoTF2023NL}.
In magnetic materials, the Onsager relation allows the unconventional
AHE with the planar configuration among the magnetization/magnetic
field, the Hall current and electric field named in-plane AHE (IPAHE)
\citep{LandauECM}. It originates from the Berry curvature of electrons
and is in contrast to the conventional planar Hall effect that has
the essential nature of anisotropic magnetoresistance \citep{pippard_magnetoresistance_1989,AAtaskin_NC_planar_2017,yin_PRL_planar_2019,li_PRB_planar_2023,zhong_CPB_recent_2023}.
Recently, the IPAHE and its quantized counterpart have attracted significant
attention and been theoretically investigated in various quantum materials,
such as thinfilms of magnetic/nonmagnetic topological insulators,
graphene-like systems, monolayer transitional metal oxides and van
der Waals heterostructure \citep{ZhangYP2011PRB,LiuX2013PRL,RenYF2016PRB,Zhong2017PRB,Sheng2017PRB,LiuZ2018PRL,You2019PRB,ZhangJL2019PRB,TanHX2021PRB,Cullen2021PRL,SunS2022PRB,LiZY2022PRL,CaoJ2023PRL}.
Remarkably, the IPAHE has been reported experimentally in heterodimensional
$\mathrm{VS_{2}-VS}$ superlattice \citep{zhouJD2022nature,CaoJ2023PRL}.
However, realistic magnetic materials with easily tunable in-plane
magnetization orientation that harbor IPAHE are still challenging.

Recently, a colossal angular-dependent magnetoresistance has been
reported in $\mathrm{Mn}_{3}\mathrm{Si}_{2}\mathrm{Te}_{6}$ single
crystal \citep{Seo2021Nature,NiYF2021PRB}, which a seven-order-of-magnitude
reduction occurs in ab plane resistivity and was attributed to metal-insulator
transition driven by lifting of the topological band degeneracy \citep{Seo2021Nature}
or the first-order ``melting transition'' driven by chiral orbital
currents \citep{ZhangY2022Nature}. Moreover, $\mathrm{Mn}_{3}\mathrm{Si}_{2}\mathrm{Te}_{6}$
exhibits a variety of intriguing quantum phenomena \citep{May2017PRB,LiuY2018PRB,Martinez2020APL,LiuY2021PRB,WangJ2022PRB,YeF2022PRB,Sala2022PRB,Mijin2023PRB,ZhangY2023PRB,RanC2023arxiv,Xue2023NSR,olmos2023},
such as polaronic transport and a sizable anomalous Nernst signal.
Notably, quasi-2D $\mathrm{Mn}_{3}\mathrm{Si}_{2}\mathrm{Te}_{6}$
nanosheet for ultrafast photonics have been obtained by a mechanical
exfoliation procedure \citep{LuY2021nanom}, enabling us to investigate
novel quantum states in few layers and monolayer of $\mathrm{Mn}_{3}\mathrm{Si}_{2}\mathrm{Te}_{6}$.
Furthermore, $\mathrm{Mn}_{3}\mathrm{Si}_{2}\mathrm{Te}_{6}$ exhibits
typical characteristics of soft magnetic materials with rapid response
to in-plane magnetic fields\citep{May2017PRB}. Thus, the monolayer
$\mathrm{Mn}_{3}\mathrm{Si}_{2}\mathrm{Te}_{6}$ provides a practical
material platform for the interplay between electronic topology, orbital
physics and magnetism.

In this work, we systematically investigate magnetism, electronic
structure, and intrinsic AHE of the monolayer $\mathrm{Mn}_{3}\mathrm{Si}_{2}\mathrm{Te}_{6}$.
We find that monolayer $\mathrm{Mn}_{3}\mathrm{Si}_{2}\mathrm{Te}_{6}$
has no imaginary frequnecy in the phonon spectrum and displays unusual
in-plane magnetization with a tiny planar magnetic anisotropy energy
(MAE). Notably, a significant sixfold AHE is found in monolayer $\mathrm{Mn}_{3}\mathrm{Si}_{2}\mathrm{Te}_{6}$
and can be controllable by manipulating the magnetization orientation.
Moreover, the strain effects on the electronic states and the AHE
are also explored. 

\begin{figure*}
\includegraphics[width=16cm]{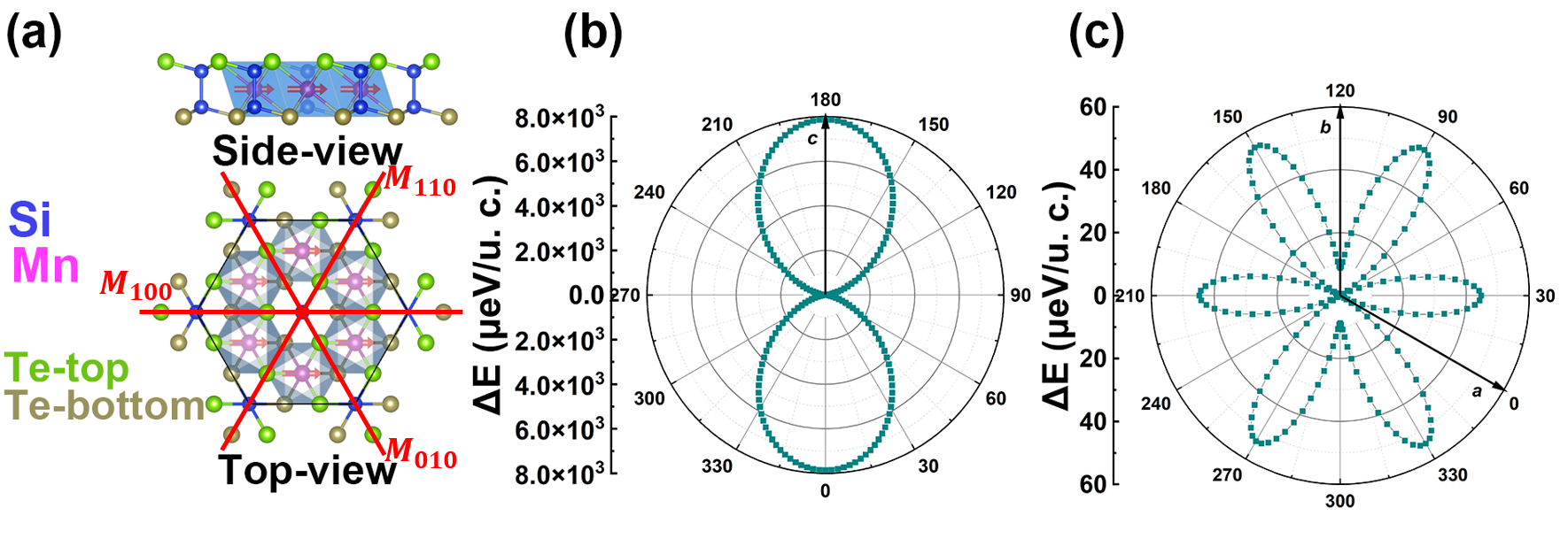}
\begin{turnpage}
\caption{\label{fig:structure} (a) The Top and side views of monolayer $\mathrm{Mn}_{3}\mathrm{Si}_{2}\mathrm{Te}_{6}$,
where the Mn, Si, Te-top and Te-bottom are represented by pink, blue,
green and brown ball, respectively. The nattier blue face represent
the distorted Te octahedral (b) The energy variation depends on the
polar angle of magnetization in the basal plane and (c) the plane
perpendicular to the a-asix of the monolayer $\mathrm{Mn}_{3}\mathrm{Si}_{2}\mathrm{Te}_{6}$.}
\end{turnpage}

\end{figure*}

\section{Structure properties}

\begin{figure*}
\includegraphics[width=16cm]{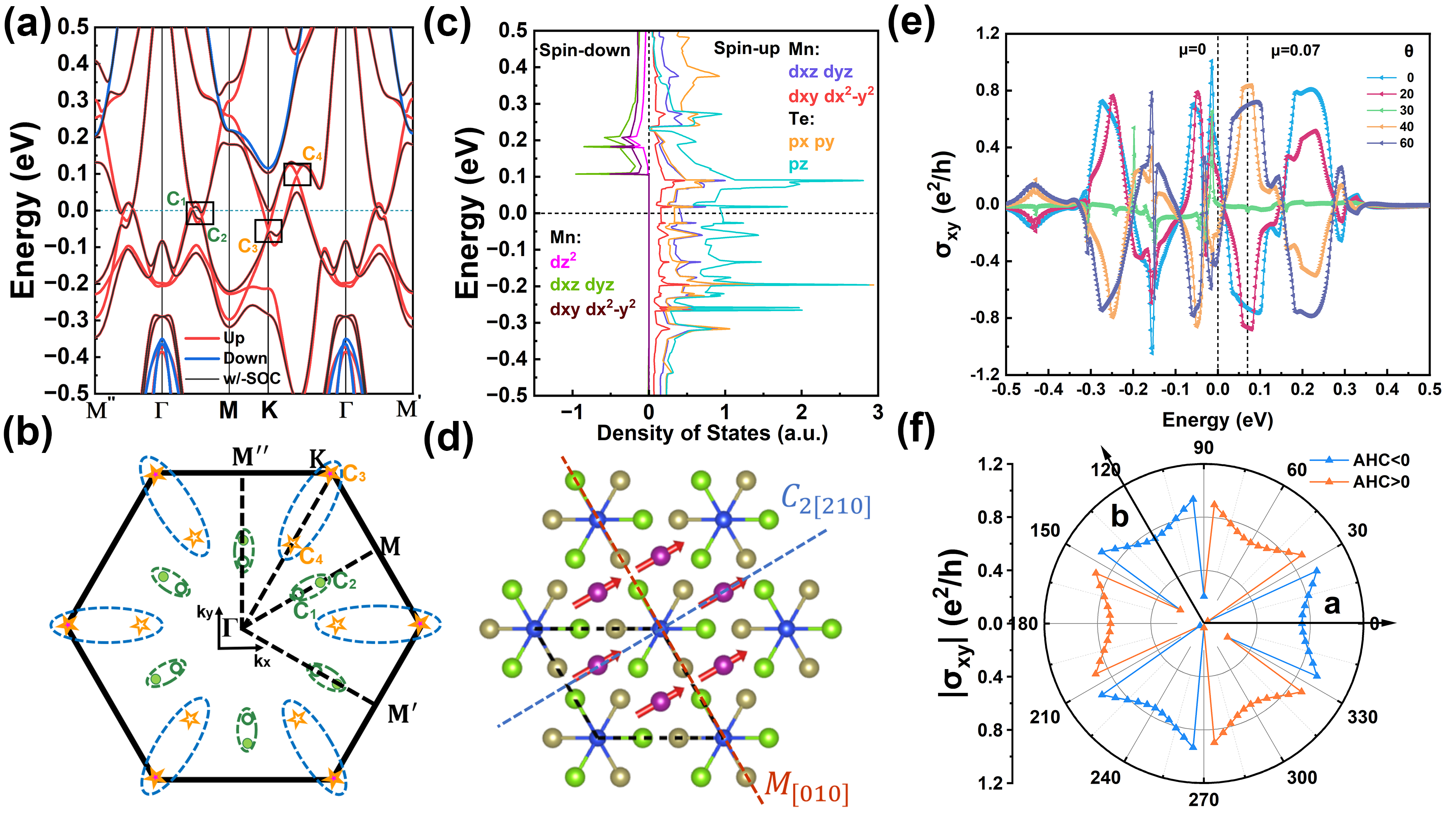}

\caption{\label{fig:eband}(a) Band structure of monolayer $\mathrm{Mn}_{3}\mathrm{Si}_{2}\mathrm{Te}_{6}$
with and without SOC. $\mathcal{C}_{n}$ labels the degenerate points.
(b) The distributions of degenerate points in the first BZ. (c) The
dominant components of PDOS of monolayer $\mathrm{Mn}_{3}\mathrm{Si}_{2}\mathrm{Te}_{6}$
without SOC. (d) The illustration of $\mathcal{C}_{2}$ and mirror
$\mathcal{M}$ symmetry at $\phi=\pi/6$. (e) The calculated AHC $\mathit{v.s.}$
Fermi energy with $\phi$ at 0°, 20°, 30°, 40° and 60°. (f) The AHC
$\mathit{v.s.}$ different in-plane magnetization with a given Fermi
energy 0.07 eV. The values represent the magnitude of AHC, and the
blue and orange colors represent negative and positive AHC, respectively.}
\end{figure*}

The ferrimagnetic compound $\mathrm{Mn}_{3}\mathrm{Si}_{2}\mathrm{Te}_{6}$
crystalizes in a layered structure with space group P$\bar{3}$1c
(No.$\:$163) \citep{RIMET1981j3m,May2017PRB}. The crystal structure
of monolayer $\mathrm{Mn}_{3}\mathrm{Si}_{2}\mathrm{Te}_{6}$ has
space group P$\bar{3}$1m (No.$\:$162) that only contains the rotational
part of P$\bar{3}$1c, as shown in Fig.$\:$\ref{fig:structure}(a).
Note that both P$\bar{3}$1m and P$\bar{3}$1c possess the same point
group D$_{3d}$. Each Mn atom is coordinated by six Te atoms forming
a distorted octahedral crystal field \citep{dresselhaus2007group}.
The Mn atoms form a simple hexagonal lattice at the 2c Wyckoff position.
Monolayer $\mathrm{Mn}_{3}\mathrm{Si}_{2}\mathrm{Te}_{6}$ includes
five atomic layers and 10 atoms per primitive hexagonal unit cell,
with the Si atom being surrounded by six Te atoms. It is noteworthy
that the Te atoms are not on the same plane and form the low-buckled
structure, which break out-of-plane mirror reflection and would be
crucial to realizing novel topological quantum states \citep{LiuCC2011PRL,Bampoulis2023PRL}.
The crystal structure of monolayer $\mathrm{Mn}_{3}\mathrm{Si}_{2}\mathrm{Te}_{6}$
has point group D$_{3d}$ with three Mirror symmetries label by $\mathcal{M}_{[0\thinspace1\thinspace0]}$,
$\mathcal{M}_{[1\thinspace0\thinspace0]}$ and $\mathcal{M}_{[1\thinspace1\thinspace0]}$,
as illustrated in Fig.$\,$\ref{fig:structure}(a). When the moments
are rotated in the x-y plane, only the mirror normal to the magnetization
direction (i.e., $\phi=(2n+1)\pi/6$, with n=0, 1, …, 5.) could be
preserved (because the magnetization is a pseudovector). For instance,
as shown in Fig.$\,$\ref{fig:eband}(d), if the magnetization direction
along $\phi=\pi/6$, the mirror $\mathcal{M}_{[0\thinspace1\thinspace0]}$
is persevered. As a result, this system has three different magnetic
space groups depending on the orientation of the moments with respect
to $\phi$. It has magnetic space group $C2^{'}/m^{'}$ and $C2/m$
at $\phi=2n\pi/6$ and $\phi=(2n+1)\pi/6$ respectively. For any other
general orientations, it has magnetic space group $P-1$ with only
two symmetry operations, i.e., $E$ (identity) and $P$ (inversion).
The lattice constants are listed in Table$\:$\ref{tab:table1}. The
structural stability is examined by the phonon spectra calculations{[}See
computational details in Appendix \ref{app.a}{]}. There are no evident
soft modes at finite q in the calculated phonon spectrum, probably
implying the mechanical stability of the monolayer $\mathrm{Mn}_{3}\mathrm{Si}_{2}\mathrm{Te}_{6}$,
as shown in Fig.$\:$\ref{fig:phonon}. Inspired by the ultrathin
$\mathrm{Mn}_{3}\mathrm{Si}_{2}\mathrm{Te}_{6}$ nanoflakes using
a simple mechanical exfoliation procedure in the previous experiment
\citep{LuY2021nanom}, it indicates that the monolayer $\mathrm{Mn}_{3}\mathrm{Si}_{2}\mathrm{Te}_{6}$
could exist stably and be prepared by the delicate mechanical exfoliation
or be growth via molecular beam epitaxy technique.

\begin{table}
\caption{\label{tab:table1} Lattice constants, thickness, bond angle of Te-Mn-Te,
and total energy $\mathit{E}$ (meV) for Néel, stripe, and zigzag
in-plane AFM configurations with Néel vector in x direction per unit
cell relative to the FM ground state.}

\begin{ruledtabular}
\begin{tabular}{ccccccc}
 & a ($\mathring{\mathrm{A}}$) & d ($\mathring{\mathrm{A}}$) & $\theta$ (°) & Néel-x & Stripe-x & Zigzag-x\tabularnewline
\hline 
Mn$_{3}$Si$_{2}$Te$_{6}$ & 6.856 & 3.559 & 86.2 & 154.6 & 23.3 & 14.2\tabularnewline
\end{tabular}
\end{ruledtabular}

\end{table}

\section{Magnetic properties}

In order to uncover the nature of the magnetic ground state of monolayer
$\mathrm{Mn}_{3}\mathrm{Si}_{2}\mathrm{Te}_{6}$, we performed the
electronic structure calculations by considering typical magnetic
configurations of Mn spin moments (see Appendix \ref{app.b})\citep{ZhangDC2020PRB,LiZY2022PRL},
i.e., ferromagnetic (FM), (ii) Néel antiferromagnetic (AFM), (iii)
stripe AFM, and (iv) zigzag AFM aligned in three directions. Table$\:$\ref{tab:table1}
lists the AFM energy of some in-plane AFM configurations with Néel
vector in x direction relative to the FM ground state in the unit
of meV. The total energies of these configurations aligned in three
directions listed in Table \ref{tab:table2} indicate that the FM
state is preferred in monolayer $\mathrm{Mn}_{3}\mathrm{Si}_{2}\mathrm{Te}_{6}$,
illustrated in Fig.$\:$\ref{fig:structure}(a). Noted that the origin
of FM order in monolayer $\mathrm{Mn}_{3}\mathrm{Si}_{2}\mathrm{Te}_{6}$
may share the same mechanism of Ruderman-Kittel-Kasuya-Yosida interaction
mediated by the itinerant carriers as the monolayer $\mathrm{Mn}\mathrm{Si}\mathrm{Te}_{3}$
with out-of-plane FM order \citep{ZhangDC2020PRB}. For the FM state,
the calculated local magnetic moments of Mn are about 3.816 $\mu_{B}$
per Mn atom, close to the experimental value in the single crystals
\citep{May2017PRB}.

We further calculate MAE to obtain the easy magnetization direction.
The MAE in different directions relative to the $\mathbf{z}$-axis
are shown in Fig.$\;$\ref{fig:structure}(b). It can be seen that
the magnetization is predominantly oriented in the basal plane with
strong anisotropy, and the dipolar energy is about 7.87 meV. In the
basal plane of the monolayer $\mathrm{Mn}_{3}\mathrm{Si}_{2}\mathrm{Te}_{6}$,
the angular dependence of the total energy (shown in Fig.$\:$\ref{fig:structure}(c))
indicates that the magnetization lies in the $\mathbf{a}$-axis or
the $60^{\circ}$ direction, with a slight degree of in-plane magnetic
anisotropy that produces only 54 $\mu\mathrm{eV}$ of dipolar energy.
The large MAE of out-of-plane magnetization and negligible difference
in in-plane magnetization imply the feasible manipulation of in-plane
magnetization orientation.

\section{Electronic structure}

Fig.$\:$\ref{fig:eband}(a) shows the band structures and density
of states (DOS) of magnetic ground state of monolayer $\mathrm{Mn}_{3}\mathrm{Si}_{2}\mathrm{Te}_{6}$
with and without spin-orbit coupling (SOC). One can see that the bands
in the spin-up channel \footnote{spin-up means that the spin direction is the same as the direction
of magnetization, while spin-down means that the spin direction is
the opposite direction of of magnetization.} are insulating with a gap of 0.46 eV, whereas only the spin-down
channel crosses the Fermi level, thus monolayer $\mathrm{Mn}_{3}\mathrm{Si}_{2}\mathrm{Te}_{6}$
has a half-metallic character. For the spin-down states, we find the
linear band crossings near the Fermi level labeled by C$_{1}$ to
C$_{4}$ in first Brillouin Zone (BZ) as shown in Fig.$\:$\ref{fig:eband}(b).
Meanwhile, we plot the dominant components of partial density of states
(PDOS) without SOC around the Fermi energy (shown in Fig.$\:$\ref{fig:eband}(c)).
It is clear that the DOS around Fermi energy are dominated by Mn $d_{xz}$,
$d_{yz}$, $d_{xy}$, $d_{x^{2}-y^{2}}$, Te $p_{x,y,z}$ orbitals
in the spin up channel. We also identify several sharp peaks that
corresponding to van Hove singularities (VHS), such as the one slightly
above the Fermi energy and the one from the nearly flat band near
the M point (about -0.3 eV). After taking account of SOC, the band
anti-crossings open small gaps, which could host nonzero Berry curvature,
implying a finite AHE in monolayer $\mathrm{Mn}_{3}\mathrm{Si}_{2}\mathrm{Te}_{6}$.

It should be noted that there are three inequivalent $\varGamma-M$
line, i.e., $\varGamma-M$, $\varGamma-M^{'}$ and $\varGamma-M^{''}$,
as shown in Fig.$\,$\ref{fig:band-theta}. For example, when the
magnetization direction is along $\phi=\pi/6$, the $\mathcal{C}$$_{2}$
operation makes the $\varGamma-M^{'}$ equivalent to the $\varGamma-M^{''}$.
Meanwhile, the mirror $\mathcal{M}_{[1\thinspace0\thinspace0]}$ operation
makes the $\varGamma-M$ equivalent to the $\varGamma-M^{'}$ for
the magnetization direction being along $\phi=\pi/2$. We also calculated
the energy dispersions along $M^{''}-\varGamma-M-K-\varGamma-M^{'}$
path for six representative orientations of the magnetic moments,
as shown in Fig.$\,$\ref{fig:eband} (b) and Appendix \ref{app.b}.
The calculated band dispersions are consistent with the symmetry analysis.
Note that the similarity of the energy dispersions for inequivalent
$\varGamma-M$ lines might come from the small planar magnetic anisotropy
energy that does cause dramatic change of energy bands except the
parts near the Fermi energy.

In fact, the two bands are degenerate protected by $\mathcal{M}_{[0\thinspace1\thinspace0]}$
for $\phi=(2n+1)\pi/6$, but become split for other orientations.
The above discussion pointed out that this material has a $type-I$
magnetic space group $C2/m$ at $\phi=(2n+1)\pi/6$. Based on the
group representation theory, the wave-vector group along $\varGamma-M$
only have one-dimensional irreducible representations\citep{dresselhaus2007group}.
Meanwhile, the two bands near crossing point belong to different irreps
$LD_{1}$ and $LD_{2}$. Therefore, the crossing states transform
according to different irreducible representations of the wave-vector
group of the high symmetry path $\varGamma-M$. Their hybridization
is prevented and a crossing at this point of intersection is symmetry
protected.

\section{Magnetization-orientation-dependent anomalous Hall effect}

Symmetry analysis provides a useful tool to understand various physical
properties of solids, such as electric conductivity tensors \citep{Grimmer1993AC,Seemann2015PRB}.
Recent work further reveals that the mirror symmetry $\mathcal{M}_{\gamma}$,
rotational symmetry $\mathcal{C}$$_{n\gamma}$, and their combination
with time reversal symmetry $\mathcal{T}$ could be effectively determine
the IPAHE in magnetic materials \citep{CaoJ2023PRL}. For the monolayer
$\mathrm{Mn}_{3}\mathrm{Si}_{2}\mathrm{Te}_{6}$, the point group
D$_{3d}$ contains three twofold rotation $\mathcal{C}_{2}$ in the
basal plane, two threefold rotation $\mathcal{C}_{3}$ perpendicular
to the basal plane and the combined symmetry with spatial inversion.
For example, the 2D Hall effect is expressed as $j_{y}=\sigma_{xy}E_{x}$.
Since the electric field is even and the electric current is odd under
the mirror reflection about the \ensuremath{y} plane ($\mathcal{M}_{x}$),
the Hall response equation becomes $j_{y}=-\sigma_{xy}E_{x}$. Consequently,
$\sigma_{xy}$ should vanish if the $\mathcal{M}_{x}$ invariant is
present. It can be also seen from the transforming of Berry curvature
under the mirror operation. The mirror operation $\mathcal{M}_{x}$
acts on the Berry curvature $\varOmega_{xy}$ as $\mathcal{M}_{x}\varOmega_{xy}(\boldsymbol{k})=-\varOmega_{xy}(-k_{x},k_{y})$.
For a system respecting $\mathcal{M}_{x}$, it leads to $\mathcal{M}_{x}\varOmega_{xy}(\boldsymbol{k})=-\varOmega_{xy}(-k_{x},k_{y})=\varOmega_{xy}(k_{x},k_{y})$.
Because the Berry curvature is an odd function of $\mathit{k}_{x}$,
the integration of Berry curvature over the whole Brillouin zone vanishes.
As known, the in-plane magnetization perpendicular to the mirror plane,
the corresponding mirror symmetry is preserved. Thus, we need to break
all in-plane mirror symmetries to realize a nonzero IPAHE as well
as its quantized version. For this material, the detailed constraint
on the IPAHE under symmetry operations are summarized in Table \ref{tab:sym-Hall}.

\begin{table}
\caption{\label{tab:sym-Hall}Constraint on the IPAHE under symmetry operations.
The label Y (N) denotes symmetry allowed (forbidden).}

\begin{ruledtabular}
\begin{tabular}{ccccccccc}
Sym. operator & $\mathcal{M}$$_{x/y}$ & $\mathcal{M}$$_{z}$ & $\mathcal{T}$$\mathcal{M}$$_{x/y}$ & $\mathcal{T}$$\mathcal{M}$$_{z}$ & $\mathcal{C}_{3z}$ & $\mathcal{T}$$\mathcal{C}_{3z}$ & $\mathcal{C}_{2}$ & $\mathcal{T}$$\mathcal{C}_{2}$\tabularnewline
\hline 
$\sigma_{Hall}$ & N & Y & Y & N & Y & N & N & Y\tabularnewline
\end{tabular}
\end{ruledtabular}

\end{table}

Fig.$\:$\ref{fig:eband}(d) shows one special orientations of magnetization
($\phi=30^{\circ}$) within the basal plane, where $\phi$ is the
angle between magnetization and the $x$ axis. In this situation,
the magnetization is perpendicular to the mirror plane $\mathcal{M}_{[0\thinspace1\thinspace0]}$,
the AHE vanishes due to the related mirror symmetry. This symmetry
analysis can be applicable to the other two mirror planes that can
be connected by $C_{3z}$ rotational operation.

It is known that, for 2D materials, the AHC can be evaluated by integrating
Berry curvature of electrons over the BZ \citep{Xiao2010RMP}. Fig.$\:$\ref{fig:eband}(e)
shows the calculated AHC relative to the Fermi level along different
magnetization orientations. One finds that, for the magnetization
aligned with the $30^{\circ}$ and $90^{\circ}$ direction and electric
current along the $x$ axis, the AHC vanish at the entire energy range\footnote{Due to the inaccuracy of numerical integral, the AHC do not equal
to zero strictly at some special energy.}. The AHCs in the directions $20^{\circ}$ and $40^{\circ}$ has almost
the same magnitude but opposite sign. The magnitudes of AHC are sensitive
to the Fermi energy, which can be enhanced by shifting the chemical
potential (e.g. $\sigma{}_{xy}=-0.86$ ($e^{2}/h$) at $\mu=0.07\:\mathrm{eV}$).

Fig.$\:$\ref{fig:eband}(f) shows the calculated AHC as a function
of the magnetization direction $\phi$ for a given chemical potential
at $\mu=0.07\:\mathrm{eV}$. As we discussed above, the symmetry restrictions
of monolayer $\mathrm{Mn}_{3}\mathrm{Si}_{2}\mathrm{Te}_{6}$ force
$\sigma_{xy}\left(\phi_{n}\right)=0$, where $\phi_{n}=(2n+1)\pi/6$
with $n=0,1,2,3,4,5$. We also found the switching of $\sigma_{xy}\left(\phi\right)$
on both sides of $\phi_{n}$, which means that one can change the
$\phi$ slightly around $\phi_{n}$ and reverse the sign of AHC. In
particular, the tiny MAE in the basal plane allows an easy switching
of AHE via a small external magnetic field to change the direction
of magnetization. The AHC with magnetization $\boldsymbol{M}$ is
marked as $\sigma_{xy}(\boldsymbol{M})$, where the angle of $\boldsymbol{M}$
to the crystallographic $\mathbf{a}$-axis is located in the region
of blue in Fig.$\:$\ref{fig:eband}(f). Next, for the opposite magnetization
$-\boldsymbol{M}$, that corresponding AHC is $\sigma_{xy}(-\boldsymbol{M})$.
The dissipationless nature of AHE implies that the Hall conductivity
is odd in magnetization or magnetic field, that is $\sigma_{xy}(\boldsymbol{M})=-\sigma_{xy}(-\boldsymbol{M})$
\citep{LandauECM}. Due to $C_{3z}$ rotational symmetry and time
reversal symmetry, one can reach the periodic AHE with $\phi$ that
disappears at special directions $\phi_{n}$. These distinctive features
of IPAHE here could be detected in the four- or six-Hall setup in
conventional transport experiments.

\begin{figure}
\includegraphics[width=8.5cm]{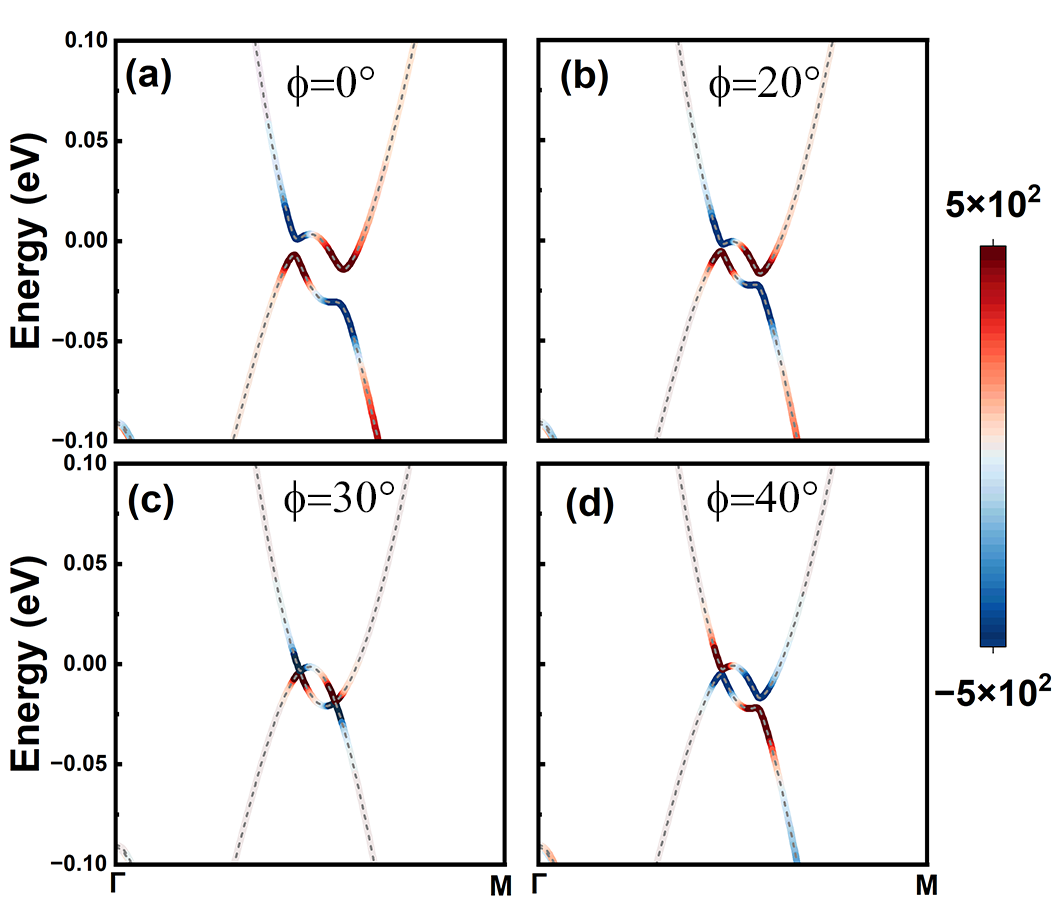}

\caption{\label{fig:Band-resolved-berry} Band resolved Berry curvature distribution
for some selected magnetization orientations in 0° (a), 20° (b), 30°
(c), and 40° (d). In the direction of $\phi=30^{\circ}$, the closure
of the energy gaps in the $\Gamma$-M path leads to AHC vanishing.
As the direction changes from 30° to 0°, the gaps increase gradually,
the AHC achieves its maximal value near 30°. For the directions (
$\phi=20^{\circ}$) and ($\phi=40^{\circ}$), the Berry curvatures
are opposite. }
\end{figure}

In order to better understand the AHE, we calculate the obtained band-resolved
Berry curvature distribution and depict the results in Fig.$\;$\ref{fig:Band-resolved-berry}{[}The
whole high symmetry path shown in Fig.$\:\ref{fig:Band-BC}${]}. It
is clear that the splitting of the doubly degenerate bands generates
a nonzero Berry curvature near the gap opening area (hot-spots), leading
to a finite AHE at $\phi=0{^\circ}$. The AHC disappears at $\phi=30^{\circ}$
due to the opposite Berry curvature contributed by the up and down
bands near the crossing point, which is consistent with the symmetry
constraint. In addition, for two angles $\phi_{1}$ and $\phi_{2}$
that are symmetric relative to the critical angles (e.g. $\phi_{1}=20^{\circ}$
and $\phi_{2}=40^{\circ}$) and break the mirror symmetry, both have
the same magnitude of Berry curvature with opposite signs resulted
from opposite-sign Rashba SOC mass term, leading to the opposite AHC.
It can be understood by the band inversion mechanism associated with
a sign reversal of energy gaps, such as near the $C_{1}$ and $C_{2}$
points \citep{Haldane1988PRL}. In addition, the Berry curvature dipole
and the resulting nonlinear Hall effect are allowed in $\mathrm{Mn}_{3}\mathrm{Si}_{2}\mathrm{Te}_{6}$
\citep{sodemann_quantum_2015,ortix_nonlinear_2021,du_quantum_2021},
deserving further discussions.

\begin{figure}
\includegraphics[width=8.3cm]{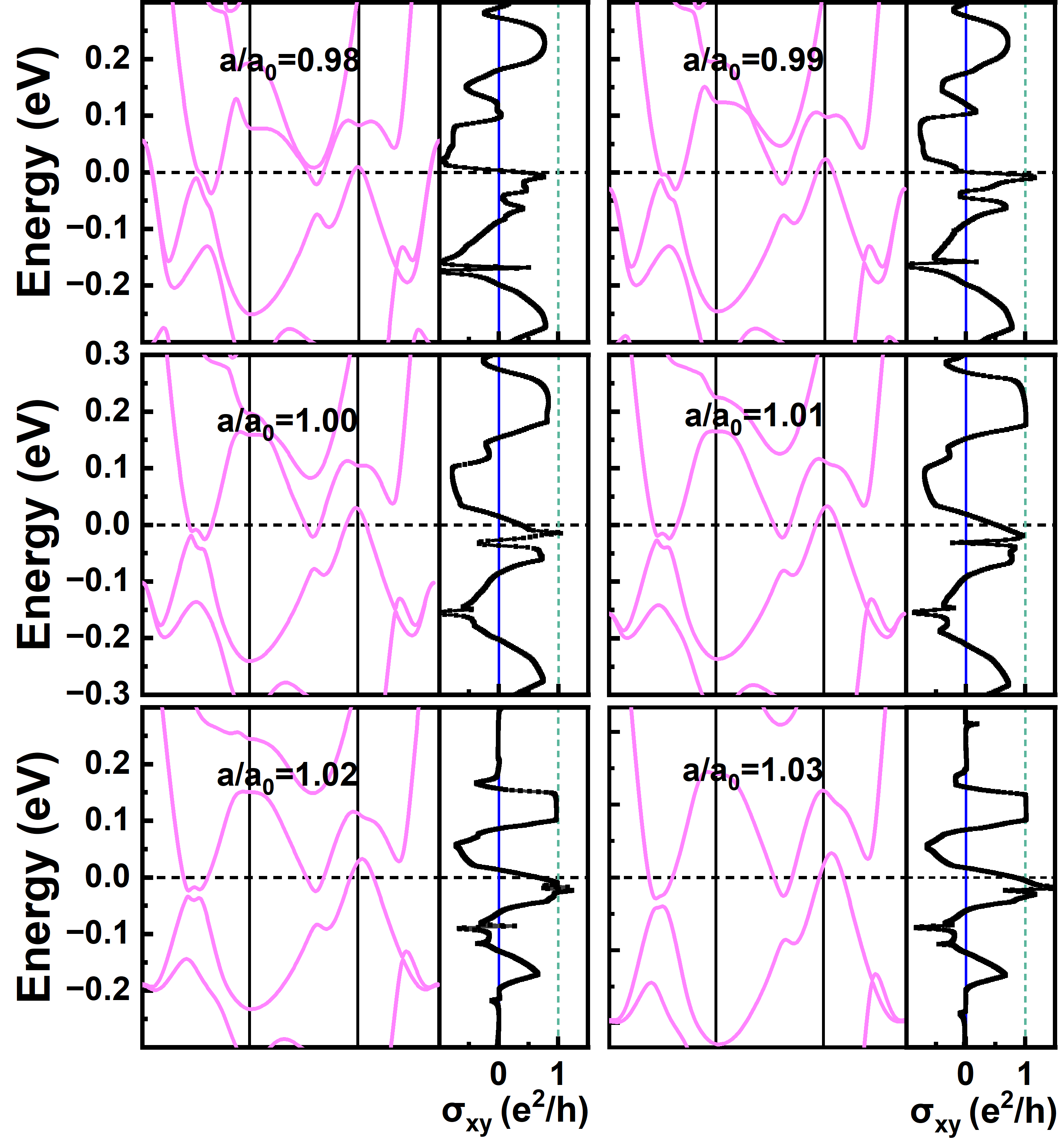}

\caption{\label{fig:AHE-band-mono} The calculated energy bands with magnetization
orientations at $\phi=0{^\circ}$ along the $\Gamma$-M-K-$\Gamma$
line and AHE for selected in-plane strains (a) $\mathrm{a/a}_{0}=0.98$,
(b) $\mathrm{a/a}_{0}=0.99$, (c) $\mathrm{a/a}_{0}=1.01$ and (d)
$\mathrm{a/a}_{0}=1.02$. $\mathrm{a/a}_{0}$ represent modification
of the lattice constant relative to the initial one $\mathrm{a}_{0}$.}
\end{figure}

Pressure or strain serves as a powerful control knob for tuning the
physical properties of solids via altering the crystal constants.
Stimulated by recent experiments of that pressure could dramatically
modify the electronic states and the novel colossal magnetoresistance
in bulk $\mathrm{Mn}_{3}\mathrm{Si}_{2}\mathrm{Te}_{6}$ \citep{WangJ2022PRB},
we thus would like to investigate the strain effect on the electronic
states and the corresponding AHE in monolayer $\mathrm{Mn}_{3}\mathrm{Si}_{2}\mathrm{Te}_{6}$
and present the calculated results in Fig.$\:$\ref{fig:AHE-band-mono}.
One can see that the energy bands near the Fermi level hosting the
Berry curvature hot-sports almost remain unchanged. Interestingly,
the moderate tensional strain ($\mathrm{a/a}_{0}=1.02$) greatly enlarges
the band separation ($0.15-0.20$ eV) along the M-K direction, leading
to a near quantized Hall conductivity {[}Fig.$\:$\ref{fig:AHE-band-mono}(d){]},
which may be observed through electric gating or doping. In comparison,
we also calculate the behaviors of energy bands and AHE versus $\mathbf{c}$-asix
strains in $\mathrm{Mn}_{3}\mathrm{Si}_{2}\mathrm{Te}_{6}$ crystal,
as shown in Fig.$\:\ref{fig:bulkAHC}$ and Fig.$\,$\ref{fig:bulk-strain}.
We find a pressure-induced semiconductor-metal transition around ($\mathrm{c/c}_{0}=1.02$),
which is consistent with recent experiments \citep{WangJ2022PRB}.
Second, the large IPAHE $\sigma_{xy}$ and $\sigma_{yz}$ appear in
the broad energy region apart from the Fermi energy ($0.30-0.45$
eV). Note that the component $\sigma_{zx}$ vanishes identically due
to the mirror symmetry $\mathcal{M}$$_{xz}$ \citep{CaoJ2023PRL,Kurumaji2023PRR}.

\section{conclusions}

Based on systematic first-principles calculations and symmetry analysis,
we predicted monolayer $\mathrm{Mn}_{3}\mathrm{Si}_{2}\mathrm{Te}_{6}$
is a stable FM material with in-plane collinear magnetization and
no imaginary phonon frequency. Meanwhile, a large sixfold IPAHE appears
in monolayer $\mathrm{Mn}_{3}\mathrm{Si}_{2}\mathrm{Te}_{6}$ and
exhibits sign reversal by manipulating magnetization orientation due
to its weak in-plane MAE. Additionally, moderate strains could effectively
change the energy bands and noticeably affect the distribution of
AHE with regard to the Fermi energy. Thus, monolayer $\mathrm{Mn}_{3}\mathrm{Si}_{2}\mathrm{Te}_{6}$
provides an ideal platform for achieving IPAHE and further investigating
the interplay among magnetism, electronic topology and exotic orbital
physics. 
\begin{acknowledgments}
\textcolor{black}{The authors acknowledge Jin Cao, Ruichun Xiao, Run
Yang and Zeying Zhang for useful discussions. This work was supported
by the National Natural Science Foundation of China under Grants (No.
U2032164 and No. 12174394)} and by the High Magnetic Field Laboratory
of Anhui Province and the HFIPS Director’s Fund (Grant Nos. YZJJQY202304
and BJPY2023B05) as well as by Chinese Academy of Sciences under contract
No. JZHKYPT-2021-08.
\end{acknowledgments}

\appendix

\section{computational details\label{app.a}}

\begin{figure}[H]
\includegraphics[width=8cm]{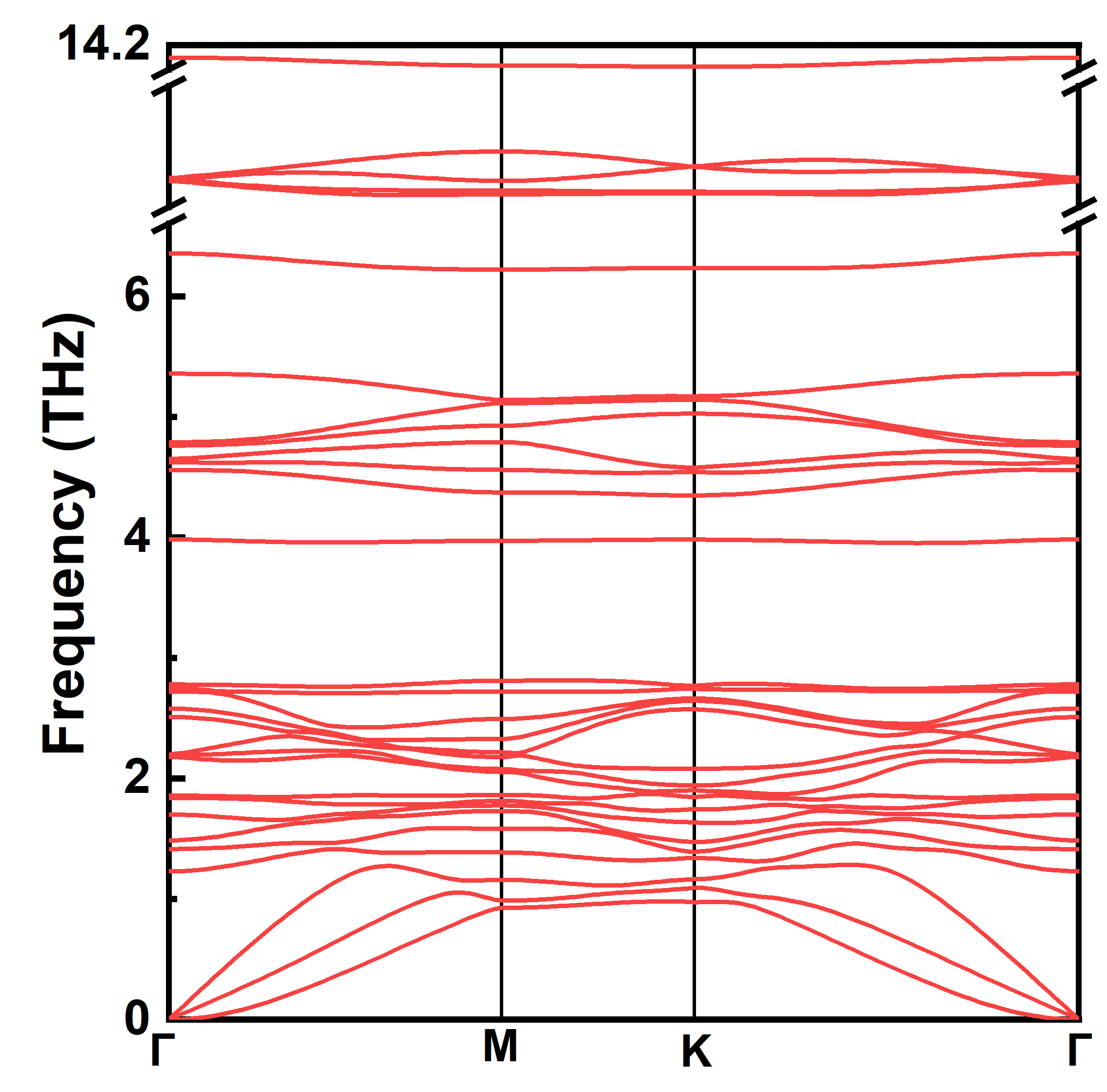}

\caption{\label{fig:phonon}The calculated phonon dispersion of monolayer $\mathrm{Mn}_{3}\mathrm{Si}_{2}\mathrm{Te}_{6}$}
\end{figure}

All DFT \citep{KohnSham1965PR,Hohenberg1964PR} calculations with
(w/) and without (w/t) spin-orbit coupling (SOC) were performed with
the Perdew-Burke-Ernzerhof (PBE) exchange-correlation functional \citep{Perdew1996PRL}
using a plane-wave basis set and projector-augmented wave method \citep{Blochl1994PRB},
as implemented in the Vienna Ab initio Simulation Package( VASP )
\citep{Kresse1996PRB}. A plane-wave basis set with a kinetic energy
cutoff of 350 eV is considered while performing first-principles calculations.
Furthermore, we have used a $\varGamma$-centered Monkhorst-Pack (7\texttimes 7\texttimes 1)
k-point mesh for the BZ sampling and Gaussian smearing of 0.05 eV.
Previous study have shown that the GGA itself provides excellent agreement
with both the magnitude of the Curie temperature and the Curie-Weiss
$\Theta$ obtained from the exchange constants \citep{May2017PRB},
so Hubbard U is not considered in our calculations. A 3\texttimes 3\texttimes 1
supercell with 3\texttimes 3\texttimes 1 k-mesh is used to obtain
phonon dispersion via Density Functional Perturbation Theory (DFPT).
The absence of imaginary phonon frequency suggests the dynamical stability
of monolayer $\mathrm{Mn}_{3}\mathrm{Si}_{2}\mathrm{Te}_{6}$, as
shown in Fig.$\:$\ref{fig:phonon}. To explore the nontrivial band
topology and the intrinsic AHE, the tight-binding Hamiltonian was
constructed with the maximally localized Wannier functions \citep{Marzari2012RMP,Marzari1997PRB,Souza2001PRB}
to reproduce closely the band structure including SOC within \textpm 1
eV of the $E_{F}$ with Mn s-d, Si s-p, and Te p orbitals. 

\section{magnetic ground state\label{app.b}}

\begin{figure}[H]
\includegraphics[width=8.5cm]{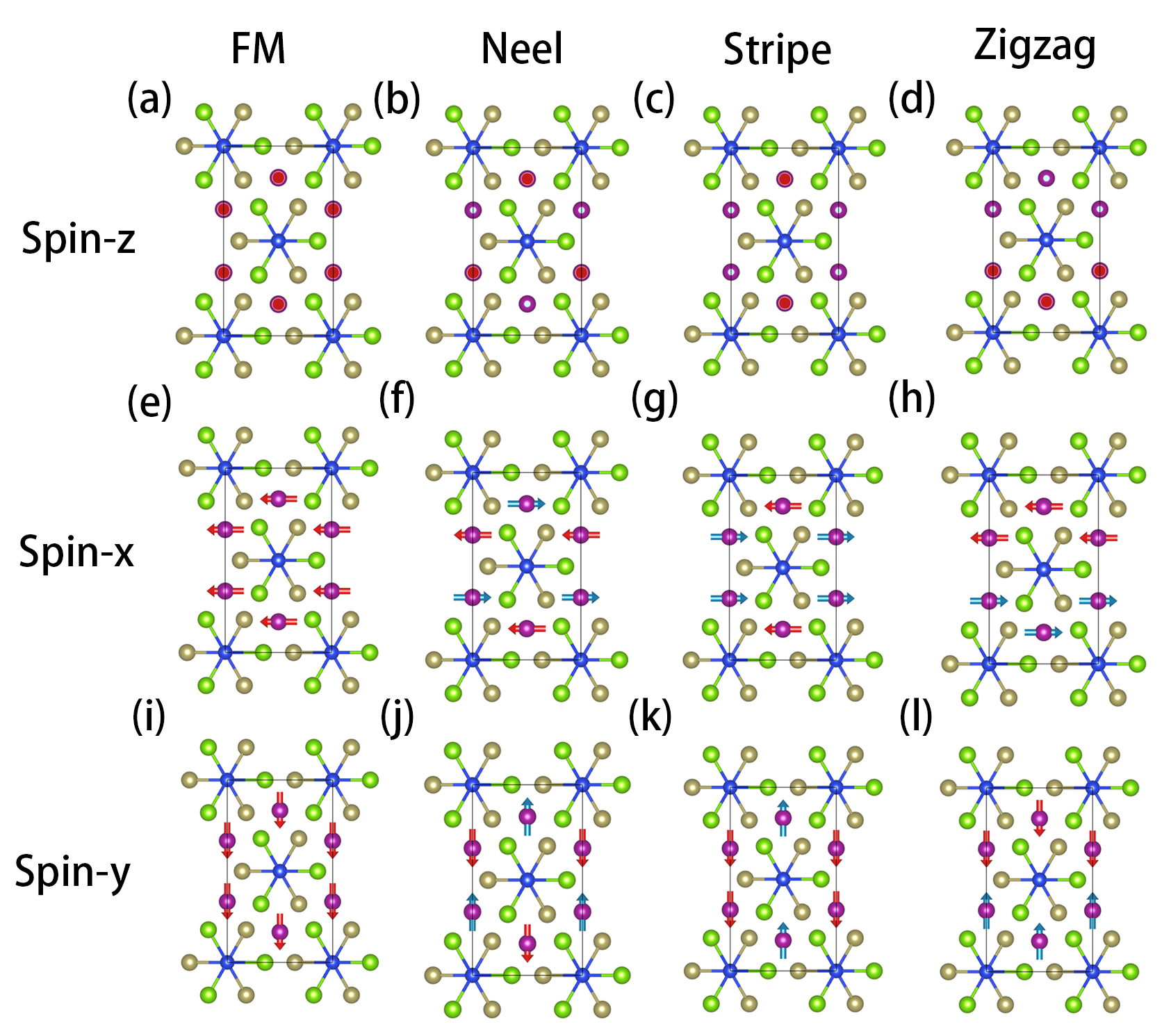}

\caption{\label{fig:configurations}FM and three AFM configurations with Néel
in x, y, and z directions corresponding to Table$\:$\ref{tab:table2}.}
\end{figure}

To obtain the magnetic ground state of monolayer $\mathrm{Mn}_{3}\mathrm{Si}_{2}\mathrm{Te}_{6}$,
we considered ferromagnetic and three antiferromagnetic configurations
with x, y, and z directions, respectively, as shown in Fig.$\:$\ref{fig:configurations}.
These calculated energy relative to the ferromagnetic ground state
were listed in Table$\:$\ref{tab:table2}. From the calculations,
magnetization prefer to lie in basal plane. Since the tiny difference
between spin-x and spin-y, detailed calculations are needed to explore
the in-plane magnetic anistropic energy(see Fig.$\,$\ref{fig:structure}
in the main text).

\begin{table}[H]
\caption{\label{tab:table2} Total energy $\mathit{E}$ (meV) for FM, Néel,
stripe, and zigzag antiferromagnetic configurations with Néel vector
in x direction per unit cell. Where the zero is set as the energy
of FM configuration with spin in x direction.}

\begin{ruledtabular}
\begin{tabular}{ccccc}
 & FM & Néel & Stripe & Zigzag\tabularnewline
\hline 
Spin-z & 18.3 & 173.7 & 42.5 & 304.2\tabularnewline
Spin-x & 0 & 183.0 & 32.5 & 297.3\tabularnewline
Spin-y & 0.0 & 184.4 & 34.7 & 263.4\tabularnewline
\end{tabular}
\end{ruledtabular}

\end{table}

\begin{figure}[H]
\includegraphics[width=8.5cm]{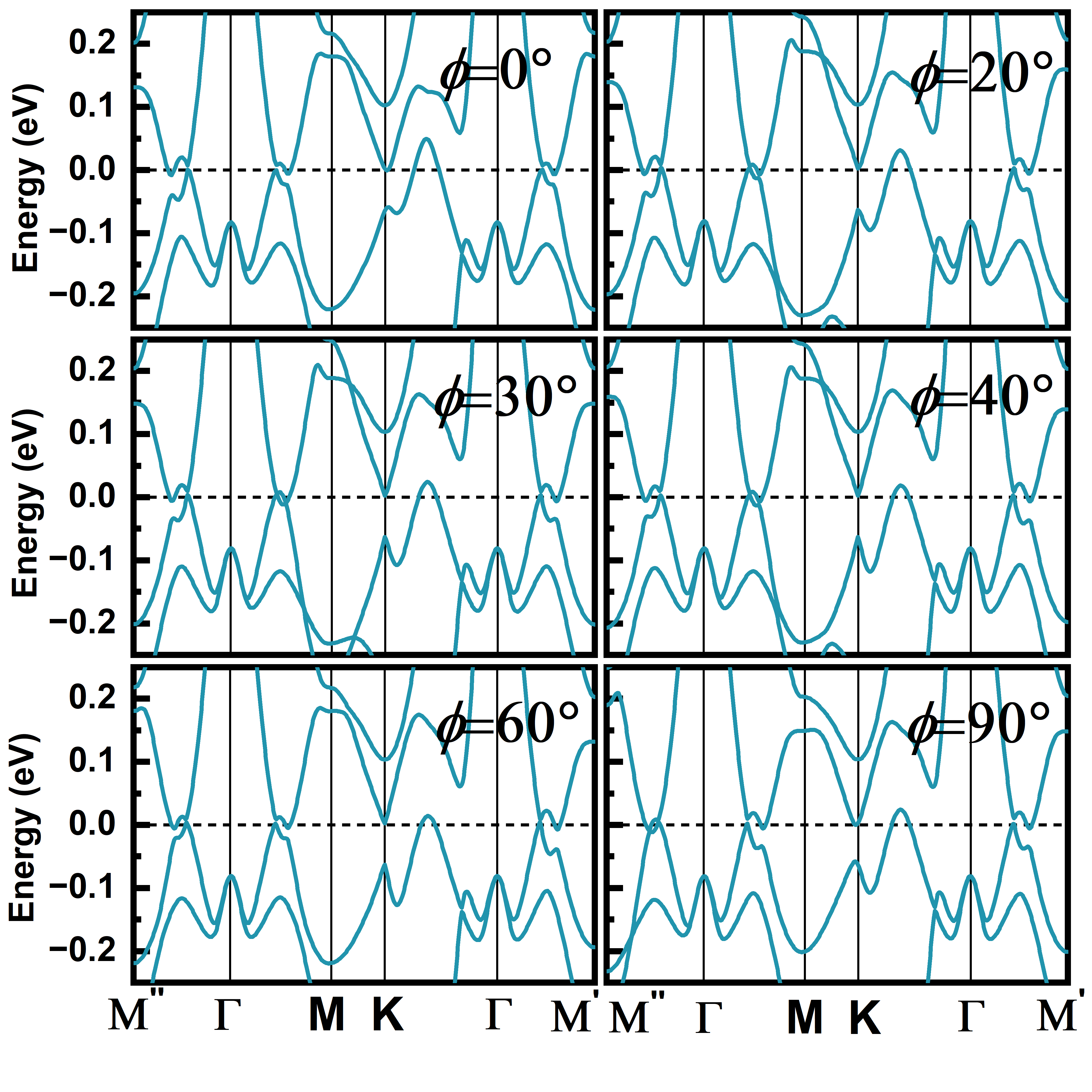}\caption{The calculated energy band of six representative orientations of the
moments. \label{fig:band-theta}}
\end{figure}

\begin{figure}[H]
\includegraphics[width=8.5cm]{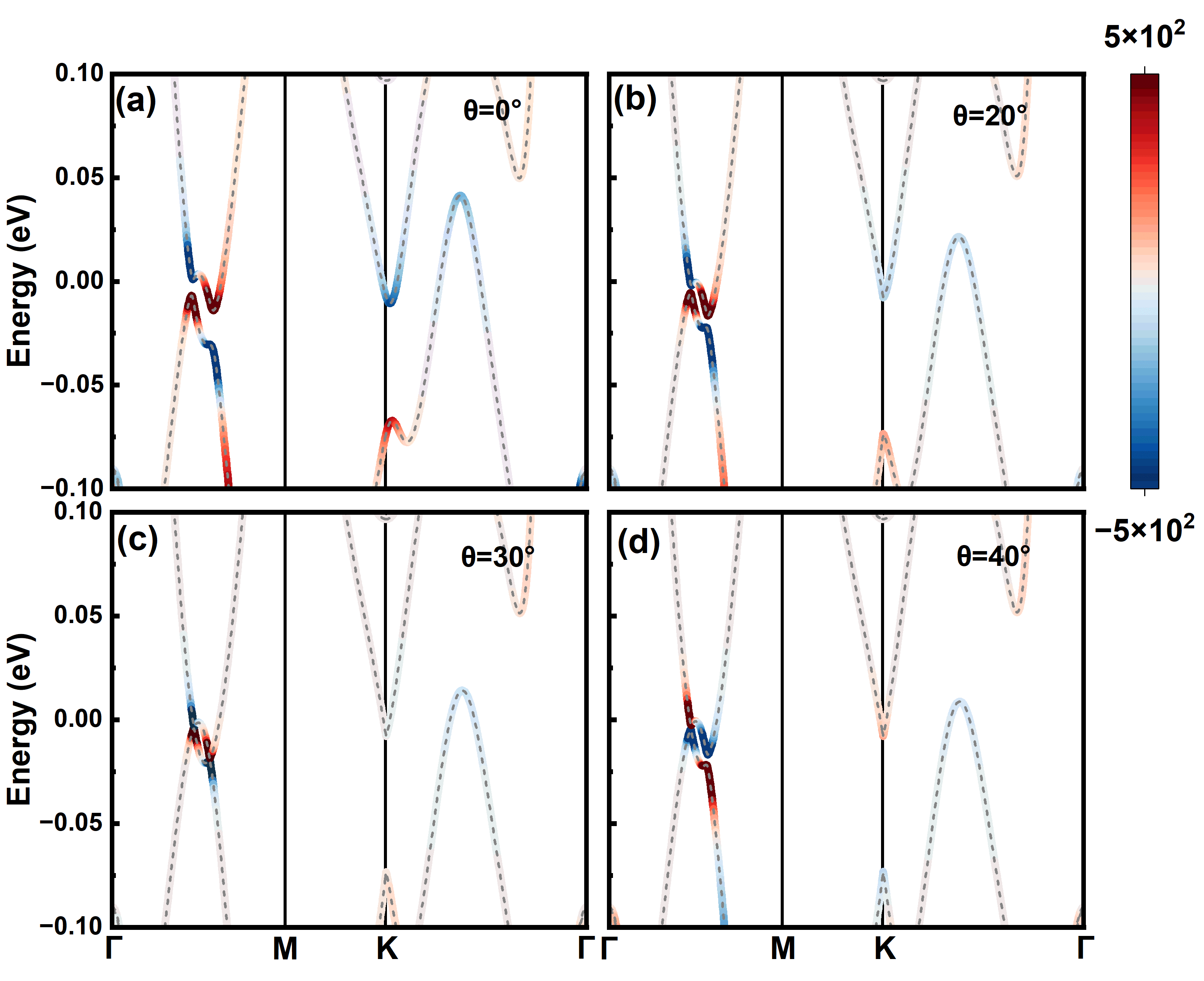}

\caption{\label{fig:Band-BC}Band resolved Berry curvature distribution for
some selected magnetization orientations in 0° (a), 20° (b), 30° (c),
and 40° (d).}
\end{figure}

\begin{figure}[H]
\includegraphics[width=8.5cm]{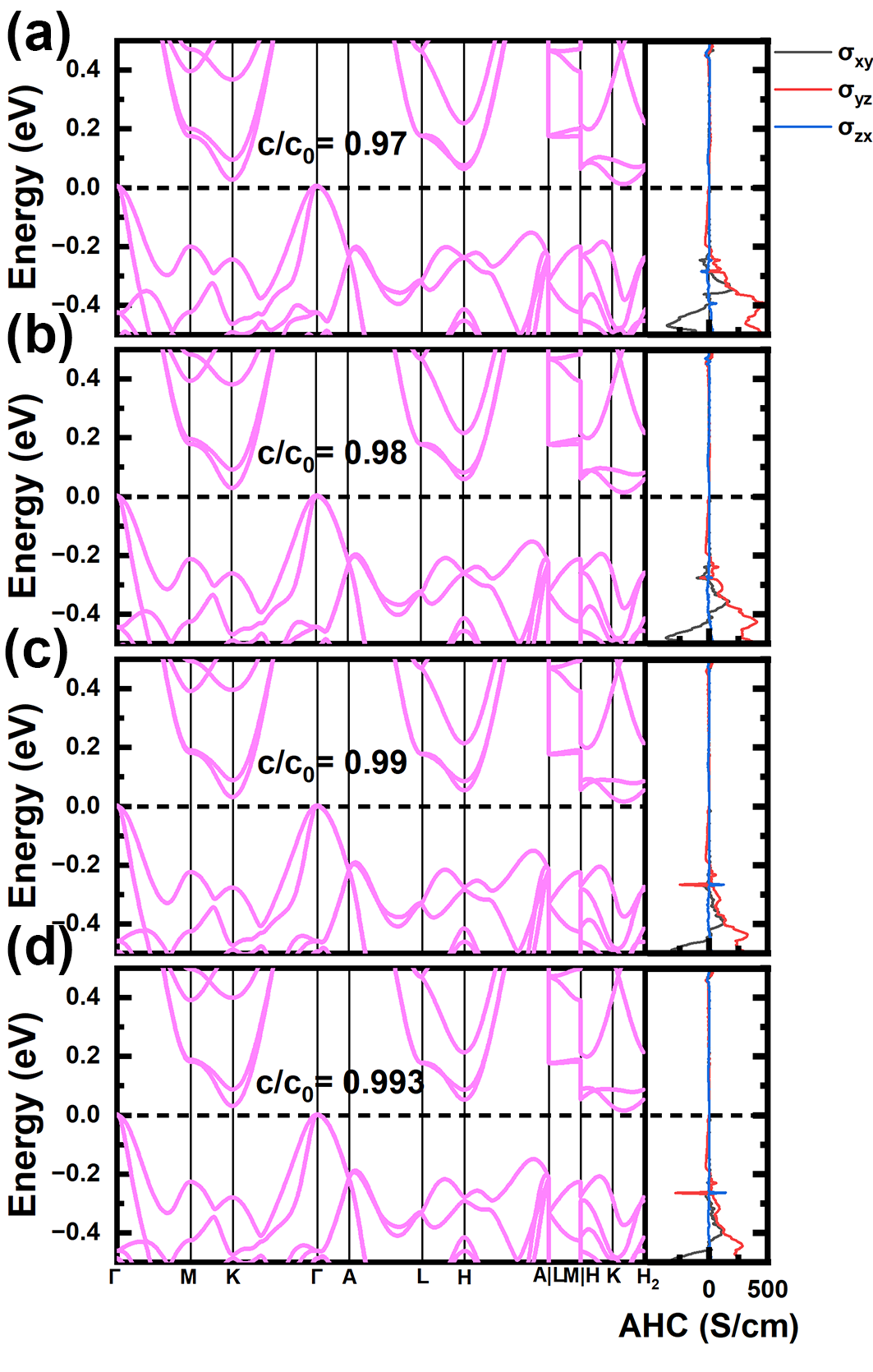}

\caption{\label{fig:bulkAHC}The AHC at different compressible strains along
$c$-axis of bulk $\mathrm{Mn}_{3}\mathrm{Si}_{2}\mathrm{Te}_{6}.$
Where the c/c$_{0}$represent the ratio between strained lattice constand
c and the initial value.}
\end{figure}

\begin{figure}[H]
\includegraphics[width=8.5cm]{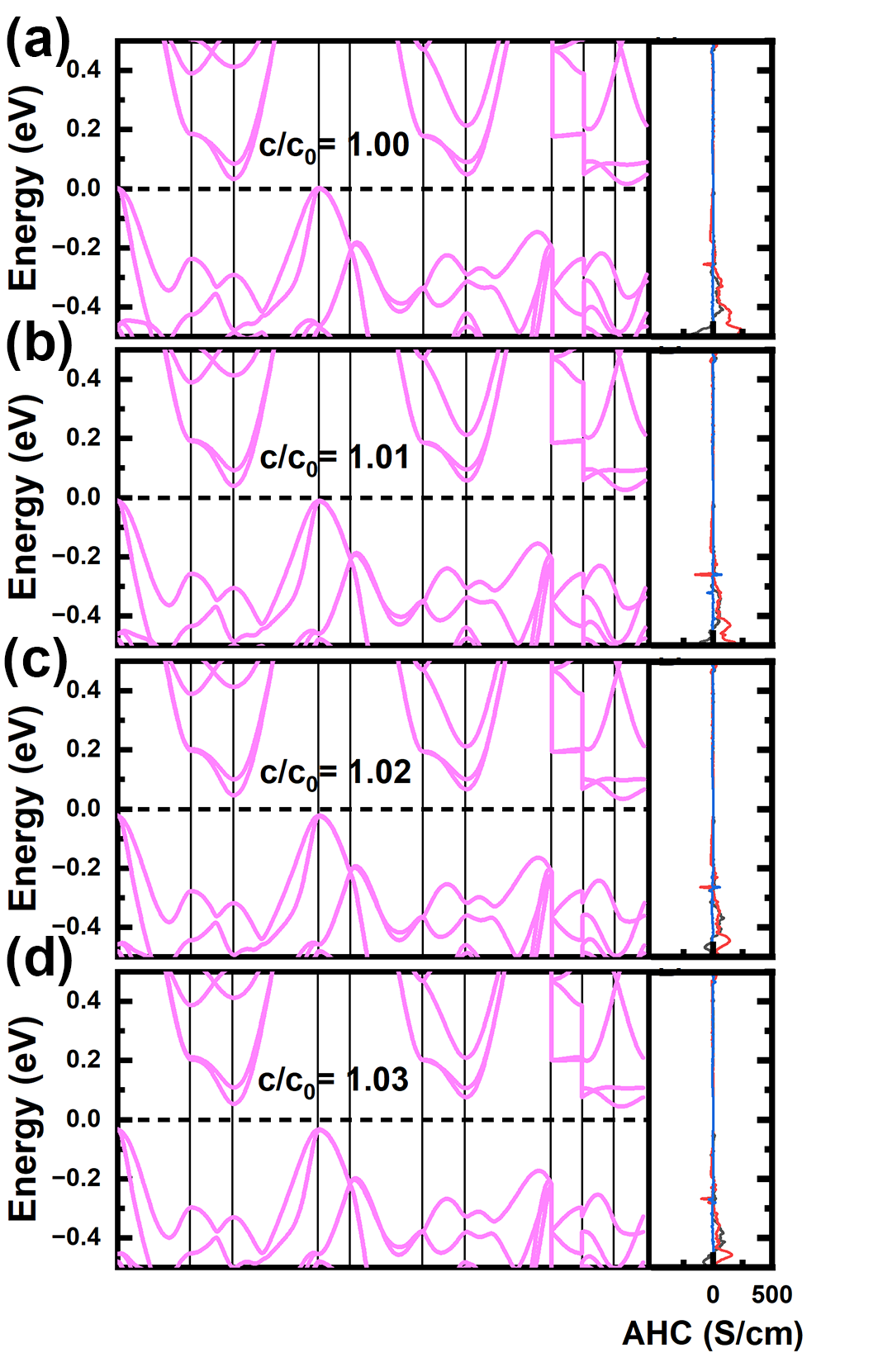}\caption{\label{fig:bulk-strain}the AHC at different tensile strains along
$c$-axis of bulk $\mathrm{Mn}_{3}\mathrm{Si}_{2}\mathrm{Te}_{6}.$
Where the c/c$_{0}$represent the ratio between strained lattice constand
c and the initial value.}
\end{figure}

\bibliographystyle{apsrev4-1}
\bibliography{ipahemst_normal}

\end{document}